\documentclass[twocolumn,prb,preprintnumbers,amsmath,amssymb,letterpaper,floatfix,showpacs]{revtex4}
\usepackage{graphicx} 
\usepackage{times}
\usepackage{gensymb } 
\usepackage{graphicx} 
\usepackage{subfigure} 
\usepackage{float} 

\begin{document}
\title{Enhancement of Superconductivity in La$_{1-x}$Sm$_{x}$O$_{0.5}$F$_{0.5}$BiS$_2$}

\author{Y. Fang$^{1,2,3}$}
\author{D. Yazici$^{2,3}$}
\author{B. D. White$^{2,3}$}
\author{M. B. Maple$^{1,2,3,}$}
\email[Corresponding Author: ]{mbmaple@ucsd.edu}

\affiliation{$^1$Materials Science and Engineering Program, University of California, San Diego, La Jolla, California 92093, USA}
\affiliation{$^2$Center for Advanced Nanoscience, University of California, San Diego, La Jolla, California 92093, USA}
\affiliation{$^3$Department of Physics, University of California, San Diego, La Jolla, California 92093, USA}

\date{\today}

\begin{abstract}
The superconducting and normal-state properties of La$_{1-x}$Sm$_{x}$O$_{0.5}$F$_{0.5}$BiS$_{2}$ (0.1 $\leqslant$ $x$ $\leqslant$ 0.9) have been studied via electrical resistivity, magnetic susceptibility, and specific heat measurements. By using suitable synthesis conditions, Sm exhibits considerable solubility into the CeOBiS$_{2}$-type LaO$_{0.5}$F$_{0.5}$BiS$_{2}$ lattice. In addition to a considerable enhancement of the superconducting volume fraction, it is found that the superconducting transition temperature \textit{T$_{c}$} is dramatically enhanced with increasing Sm concentration to 5.4 K at $x$ = 0.8. These results suggest that \textit{T$_{c}$} for SmO$_{0.5}$F$_{0.5}$BiS$_{2}$ could be as high as $\sim$6.2 K and comparably high \textit{T$_{c}$} values might also be obtained in \textit{Ln}O$_{0.5}$F$_{0.5}$BiS$_{2}$ (\textit{Ln} = Eu - Tm) if these compounds can be synthesized.
\end{abstract}
\pacs{74.25.F-, 74.25.Dw, 74.62.Bf}

\maketitle


\section{INTRODUCTION}
Since the discovery of superconductivity in Bi$_{4}$O$_{4}$S$_{3}$,\cite{1,2} a tremendous amount of effort has been made to synthesize new superconducting materials with BiS$_{2}$-layers.  Through fluorine substitution for oxygen, the compounds \textit{Ln}O$_{1-x}$F$_{x}$BiS$_{2}$ (\textit{Ln} = La, Ce, Pr, Nd, Yb) were soon reported to have superconducting transition temperatures, \textit{T$_{c}$}, ranging from 2 to 10 K.\cite{6,7,8,9,10,11,12,13,14,15,16,Inho} Superconductivity can also be induced in LaOBiS$_{2}$ via substitution of tetravalent elements, such as Th$^{4+}$, Hf$^{4+}$, Zr$^{4+}$, and Ti$^{4+}$, for trivalent \textit{Ln}$^{3+}$.\cite{14} Very recently, bulk superconductivity was observed in La substituted SrFBiS$_{2}$.\cite{Sr1} These compounds form in a tetragonal  structure with  space group \textit{P4/nmm}, composed of alternate stacking of double superconducting BiS$_{2}$ layers and blocking \textit{Ln}O or SrF layers.\cite{7,8,9,11,13} Thus, there exists significant phase space to design and synthesize analogous superconductors by changing the chemical environment of the blocking layers or modifying the superconducting layers.

The \textit{T$_{c}$} values for samples of the superconducting compounds \textit{Ln}O$_{1-x}$F$_{x}$BiS$_{2}$, prepared at ambient pressure, increase with increasing atomic number for \textit{Ln} = La - Nd.\cite{11,16} Non-superconducting samples of \textit{Ln}BiOS$_{2}$ (\textit{Ln} = La, Ce, Pr, Nd, Sm, Gd, Dy, Yb) were successfully synthesized decades ago;\cite{Ln} however, attempts to prepare fluorine-substituted samples of \textit{Ln}BiO$_{1-x}$F$_{x}$S$_{2}$ for \textit{Ln} = Sm - Tm, which could potentially exhibit superconductivity, have been unsuccessful. Since the highest \textit{T$_{c}$} in as-grown samples of LaO$_{1-x}$F$_{x}$BiS$_{2}$ is $\sim$2.8 K for $x$ = 0.5, we felt that it would be instructive to systematically substitute Sm for La in LaO$_{0.5}$F$_{0.5}$BiS$_{2}$ in order to determine the solubility limit and to address the question of whether SmO$_{0.5}$F$_{0.5}$BiS$_{2}$ might be a superconductor.

In this paper, we study the evolution of superconductivity as well as the normal-state properties of polycrystalline samples of La$_{1-x}$Sm$_{x}$O$_{0.5}$F$_{0.5}$BiS$_{2}$ from $x$ = 0.1 to the Sm solubility limit near $x$ = 0.8. Evidence for an enhancement with $x$ of both \textit{T$_{c}$} and the volume fraction is presented. The increasing volume fraction suggests that high-quality samples of SmO$_{0.5}$F$_{0.5}$BiS$_{2}$ that exhibit bulk superconductivity could be produced if the phase could be stabilized. Performing a linear extrapolation of \textit{T$_{c}$} vs. $x$ to $x$ = 1 allowed us to estimate \textit{T$_{c}$} $\sim$6.2 K for SmO$_{0.5}$F$_{0.5}$BiS$_{2}$. The results are consistent with the trend of \textit{T$_{c}$} vs. \textit{Ln} for the reported \textit{Ln}O$_{0.5}$F$_{0.5}$BiS$_{2}$ compounds. Until the heavy lanthanide variants can be synthesized, the results reported herein for \textit{Ln} = Sm  constitute a test case for a promising approach to make a preliminary assessment of superconductivity in \textit{Ln}O$_{0.5}$F$_{0.5}$BiS$_{2}$ compounds.

\section{EXPERIMENTAL DETAILS}
High-quality La$_{1-x}$Sm$_{x}$O$_{0.5}$F$_{0.5}$BiS$_{2}$ samples were synthesized by means of solid state reaction as described elsewhere.\cite{11} Powder X-ray diffraction experiments were performed at room temperature using a Bruker D8 Discover x-ray diffractometer with Cu-K$_\alpha$ radiation. All resulting patterns were analyzed by Rietveld refinement using the GSAS+EXPGUI software package.\cite{gsas,expgui} Electrical resistivity measurements were performed by means of a standard four-wire technique using a Linear Research LR700 ac impedance bridge and a home-built probe in a liquid $^4$He Dewar from 300 K to $\sim$1.1 K. Alternating current (ac) magnetic susceptibility measurements were made down to $\sim$1.1 K in a liquid $^4$He Dewar using home-built magnetic susceptibility coils and the Linear Research LR700 impedance bridge. Direct current (dc) magnetic susceptibility measurements were carried out using a Quantum Design magnetic properties measurement system (MPMS). Specific heat measurements were performed in a Quantum Design Physical Property Measurement System (PPMS) Dynacool using a standard thermal relaxation technique.

\section{RESULTS AND DISCUSSION}

A representative XRD pattern for La$_{1-x}$Sm$_{x}$O$_{0.5}$F$_{0.5}$BiS$_{2}$ with $x$ = 0.7 is shown in Fig. 1, plotted with its refined pattern for comparison. For $x$ $\leqslant$ 0.8, the main diffraction peaks can be fitted well by the Rietveld refinement method to a CeOBiS$_{2}$-type tetragonal crystal structure with space group \textit{P4/nmm}. Possible impurities, which
seem to be independent of the nominal Sm concentration up to $x$ = 0.8, were found to be La(Sm)F$_{3}$ ($\textless$ 10 wt.\%), La(Sm)O ($\textless$ 3 wt.\%) and Bi$_{2}$S$_{3}$ ($\textless$ 1 wt.\%), resulting in a slight discrepancy between nominal and actual chemical composition of the main phase. For $x$ $\geqslant$ 0.9, samples contain a considerable amount of impurities and the parent compound SmO$_{0.5}$F$_{0.5}$BiS$_{2}$ could not be synthesized, indicating a Sm solubility limit near 80\%. The main diffraction peaks, \{102\} and \{004\}, shift with increasing $x$ (see the inset of Fig. 1), indicating a systematic change in the lattice parameters. The Sm concentration dependence of the lattice parameters \textit{a}, \textit{c}, and unit-cell volume \textit{V} for $x$ = 0.1 to 0.8 are summarized in Fig. 2. Although superconductivity is observed in the nominal $x$ = 0.9 sample, its lattice parameters are not plotted here because of appreciable amount of impurities that make XRD analysis unreliable. As the Sm concentration increases from $x$ = 0.1 to 0.8, the \textit{a} axis decreases continuously, while, the \textit{c} axis increases, leading to a decrease in unit-cell volume of $\sim$3\%. Extrapolation of the unit-cell volume linearly to $x$ = 1 provides an estimated volume $V$ = 212.5 \AA$^{3}$ for SmO$_{0.5}$F$_{0.5}$BiS$_{2}$ (see Fig. 2(c)), which is consistent with that of other reported \textit{Ln}O$_{1-x}$F$_{x}$BiS$_{2}$ compounds in which the \textit{Ln} ion is believed to be trivalent.\cite{11}
\begin{figure}[H]
\centering
\includegraphics[width=8.5cm]{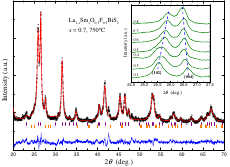}
\caption{(Color online) X-ray diffraction pattern of La$_{0.3}$Sm$_{0.7}$O$_{0.5}$F$_{0.5}$BiS$_{2}$ as a representative example. Black crosses denote the experimental data. Red and blue lines are the calculated XRD pattern and the difference between the observed and calculated patterns, respectively. Tick marks represent calculated peak positions of the main phase (purple) and LaF$_{3}$ (orange). (Inset) XRD profiles of \{102\} and \{004\} peaks of $x$ = 0.1 to 0.8. The dashed lines are guides to the eye.}
\label{FIG.1.}
\end{figure}

\begin{figure}[t]
\includegraphics[width=8.5cm]{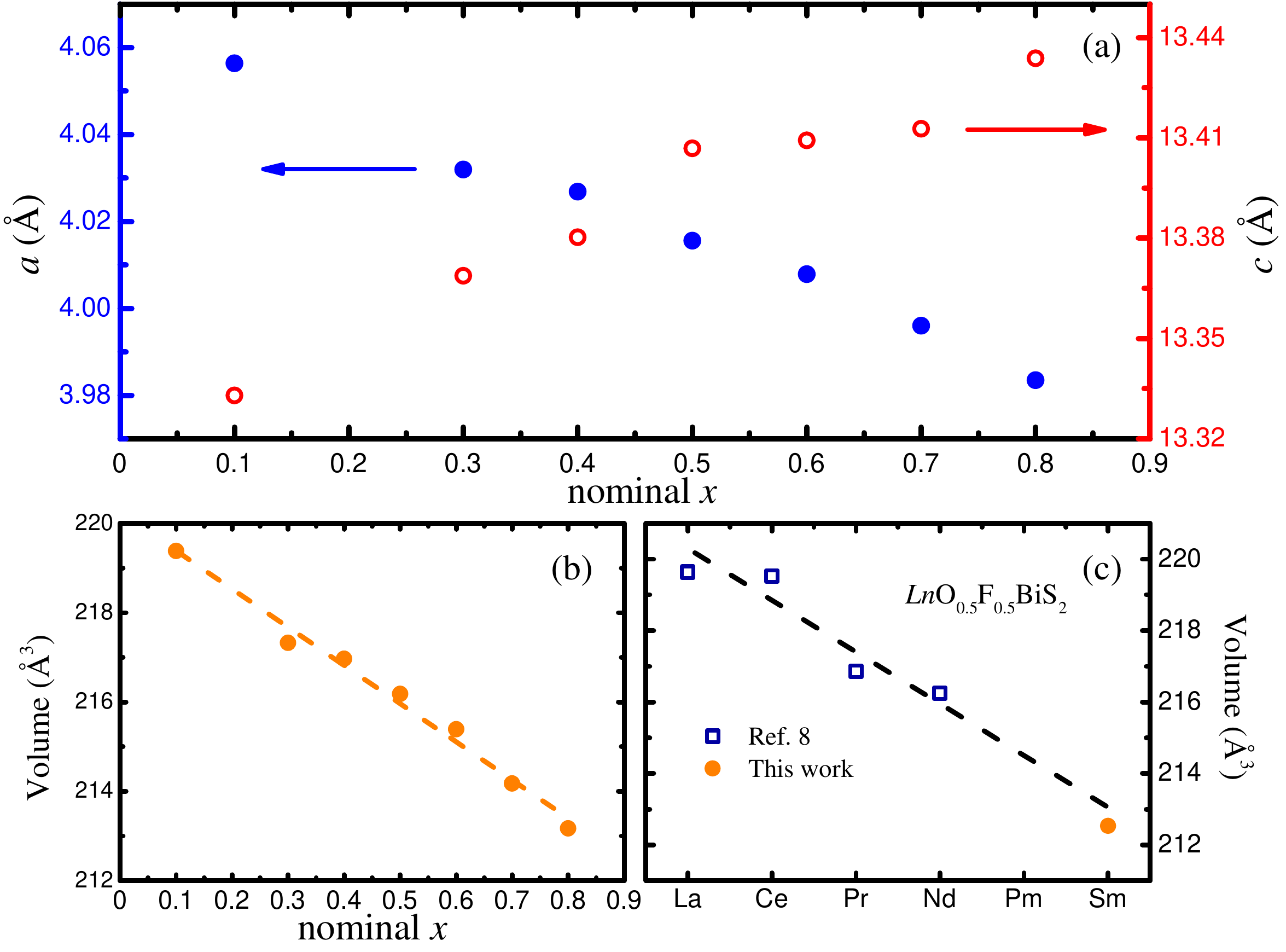}
\caption{(Color online) Dependence of (a) lattice parameters \textit{a} (left axis) and \textit{c} (right axis), and (b) unit-cell volume $V$ on nominal Sm concentration $x$. (c) $V$ for \textit{Ln}O$_{0.5}$F$_{0.5}$BiS$_{2}$ (\textit{Ln} = La, Ce, Pr, Nd) from Ref. 8 and estimated $V$ for SmO$_{0.5}$F$_{0.5}$BiS$_{2}$. Dashed lines are guides to the eye.}
\label{FIG.2.}
\end{figure}

Electrical resistivity $\rho$ vs. temperature $T$ in zero magnetic field is depicted in Fig. 3. Upon cooling, $\rho$(\textit{T}) increases until the onset of the superconducting transition for all samples, indicating semiconducting-like behavior. However, this behavior is suppressed with Sm substitution, which implies that the samples with higher Sm concentration have smaller semiconducting energy gaps compared with those with lower Sm concentration. With Sm substitution for La, the \textit{T$_{c}$} of La$_{1-x}$Sm$_{x}$O$_{0.5}$F$_{0.5}$BiS$_{2}$ gradually increases and reaches a maximum value of \textit{T$_{c, \rho}$} = 5.4 K for $x$ = 0.8 as is shown in Fig. 4. The value of \textit{T$_{c, \rho}$} is defined by the temperature where the electrical resistivity falls to 50\% of its normal-state value, and the width of the transition is characterized by identifying the temperatures where the electrical resistivity decreases to 90\% and 10\% of that value. For $x$ = 0.9, due to the presence of an appreciable amount of impurities like LaF$_{3}$, the actual chemical composition of the sample is probably quite different from the nominal composition (i.e., less fluorine). This would be expected to cause a decrease in \textit{T$_{c}$},\cite{9} which is consistent with our results. Extrapolating \textit{T$_{c, \rho}$}($x$) for $x$ $\leqslant$ 0.8 linearly to $x$ = 1 yields an estimate for the expected \textit{T$_{c}$} of SmO$_{0.5}$F$_{0.5}$BiS$_{2}$ of $\sim$6.2 K (see inset of Fig. 4), which is significantly higher than the \textit{T$_{c}$} reported for other \textit{Ln}O$_{0.5}$F$_{0.5}$BiS$_{2}$ compounds synthesized at ambient pressure.\cite{11,16}

The effects of annealing temperature on the electrical resistivity and \textit{T$_{c}$} were also investigated. An annealing temperature of 800$\celsius$ is suitable to prepare La$_{1-x}$Sm$_{x}$O$_{0.5}$F$_{0.5}$BiS$_{2}$ samples for $x$ $\leqslant$ 0.3. However, for $x$ $\geqslant$ 0.5, annealing the samples at 800$\celsius$ caused a significant increase in the amount of impurities, resulting in a large normal-state electrical resistivity and low \textit{T$_{c}$}. To reduce the concentration of these impurities, different heat-treatment temperatures were used to synthesize the samples. By decreasing the annealing temperature, it was possible to significantly enhance \textit{T$_{c}$} and decrease the normal-state electrical resistivity (see Fig. 3(b)) for the samples with high Sm concentrations. On the other hand, when the $x$ = 0.8 sample is annealed at 750$\celsius$, \textit{T$_{c}$} is very similar but the electrical resistivity is slightly lower, compared with the \textit{T$_{c}$} and resistivity values for samples annealed at 710$\celsius$. This suggests that the optimal annealing temperature for synthesizing La$_{1-x}$Sm$_{x}$O$_{0.5}$F$_{0.5}$BiS$_{2}$ samples with $x$ $\geqslant$ 0.5 is probably around 750$\celsius$.
\begin{figure}[t]
\centering
\includegraphics[width=8.5cm]{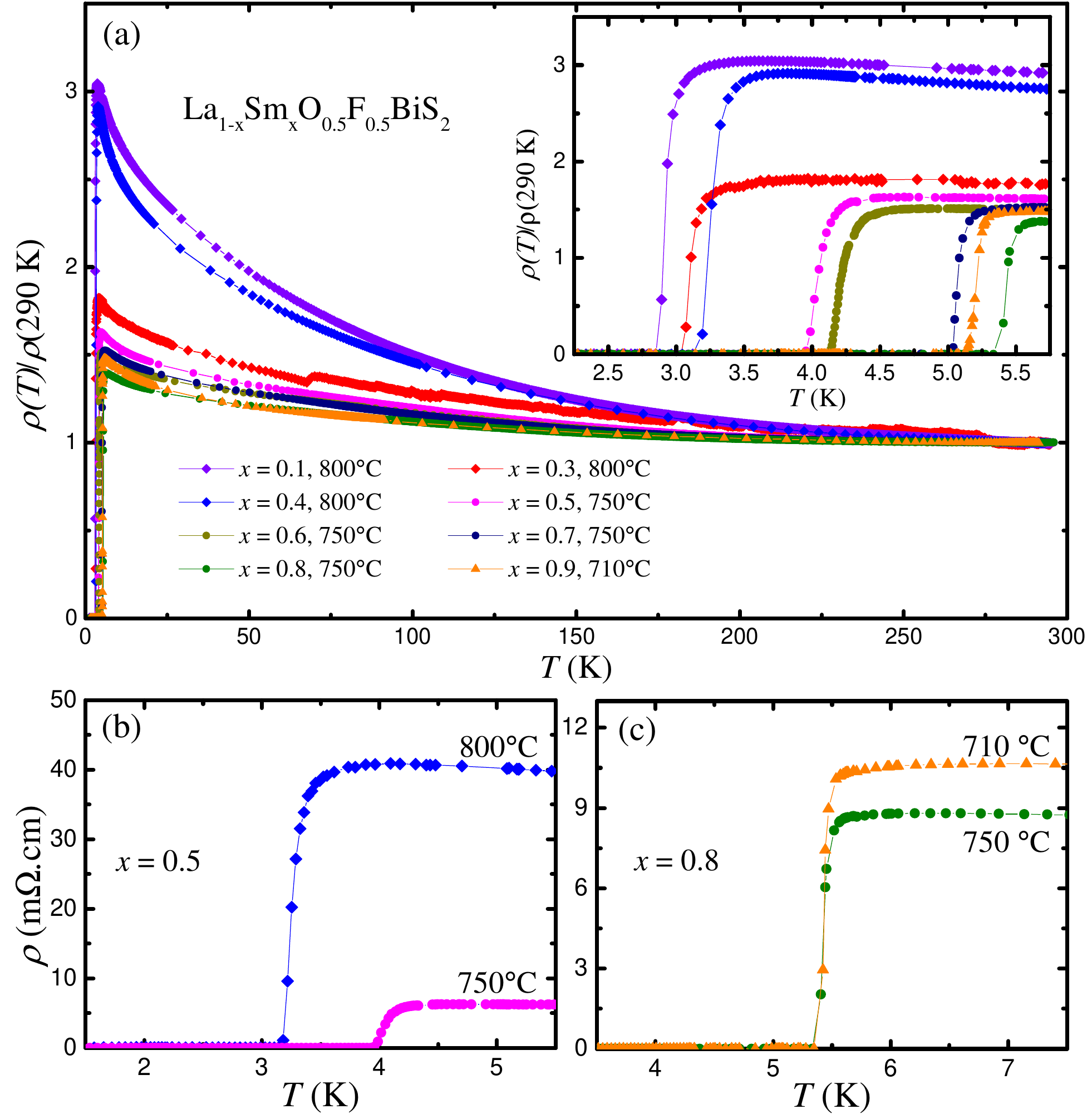}
\caption{(Color online) (a) Temperature dependence of the electrical resistivity, $\rho$(\textit{T}), normalized by its value at 290 K, $\rho$(290 K), for La$_{1-x}$Sm$_{x}$O$_{0.5}$F$_{0.5}$BiS$_{2}$. The inset displays the data in panel (a) from 2 to 6 K, emphasizing the superconducting transitions. (b) and (c) Electrical resistivity $\rho$(\textit{T}) for two samples with $x$ = 0.5 and $x$ = 0.8, respectively. The annealing temperature used for each sample is denoted.}
\label{FIG.3.}
\end{figure}

Figures 5(a) show the temperature dependence of zero-field-cooled (ZFC) and field-cooled (FC) dc magnetic susceptibility under an applied magnetic field of 10 Oe. The samples exhibit reasonable diamagnetic signals associated with the induced supercurrent during ZFC measurements, suggesting that the sample are bulk superconductors. A paramagnetic contribution to the magnetic susceptibility around \textit{T$_{c}$}, which is larger at lower external magnetic fields and for higher Sm concentration samples, was observed during both FC and ZFC measurements as shown in Fig. 5(b).  Similar features have been reported in certain copper oxide superconductors, Nb disks, MgB$_{2}$, and Pb, and are generally referred to a paramagnetic Meissner effect (PME).\cite{paramagnetic1,paramagneticnb,paramagneticnbHTSC,paramagneticnbMgB2,paramagneticPb,magnetic} However, further work needs to be done in order to determine whether the observed paramagnetic signal is associated with the PME or is related to movement of the samples in an inhomogeneous external magnetic field in the MPMS system.\cite{Artifacts1,Artifacts2} A jump from negative to positive magnetic susceptibility during ZFC measurements in the data for $x$ = 0.7, 0.8 is an instrumental artifact resulting from a brief loss of the temperature control near the boiling point of $^4$He, during which the temperature will suddenly increase above \textit{T$_{c}$} and then slowly return to the set point. This induces extra irreversible magnetic flux penetration during ZFC measurements in the samples with \textit{T$_{c}$} $\textgreater$ 4.4 K. AC magnetic susceptibility data for selected samples with $x$ = 0.1, 0.5, 0.7, and 0.9 are plotted in Figs. 5(c) and (d). The smooth transitions in both ac and dc magnetic susceptibility data imply there is probably only one phase that contributes to the observed bulk shielding signal. No evidence of a structural phase transition induced by Sm substitution was observed.\cite{17,18,19}
\begin{figure}[t]
\includegraphics[width=8.5cm]{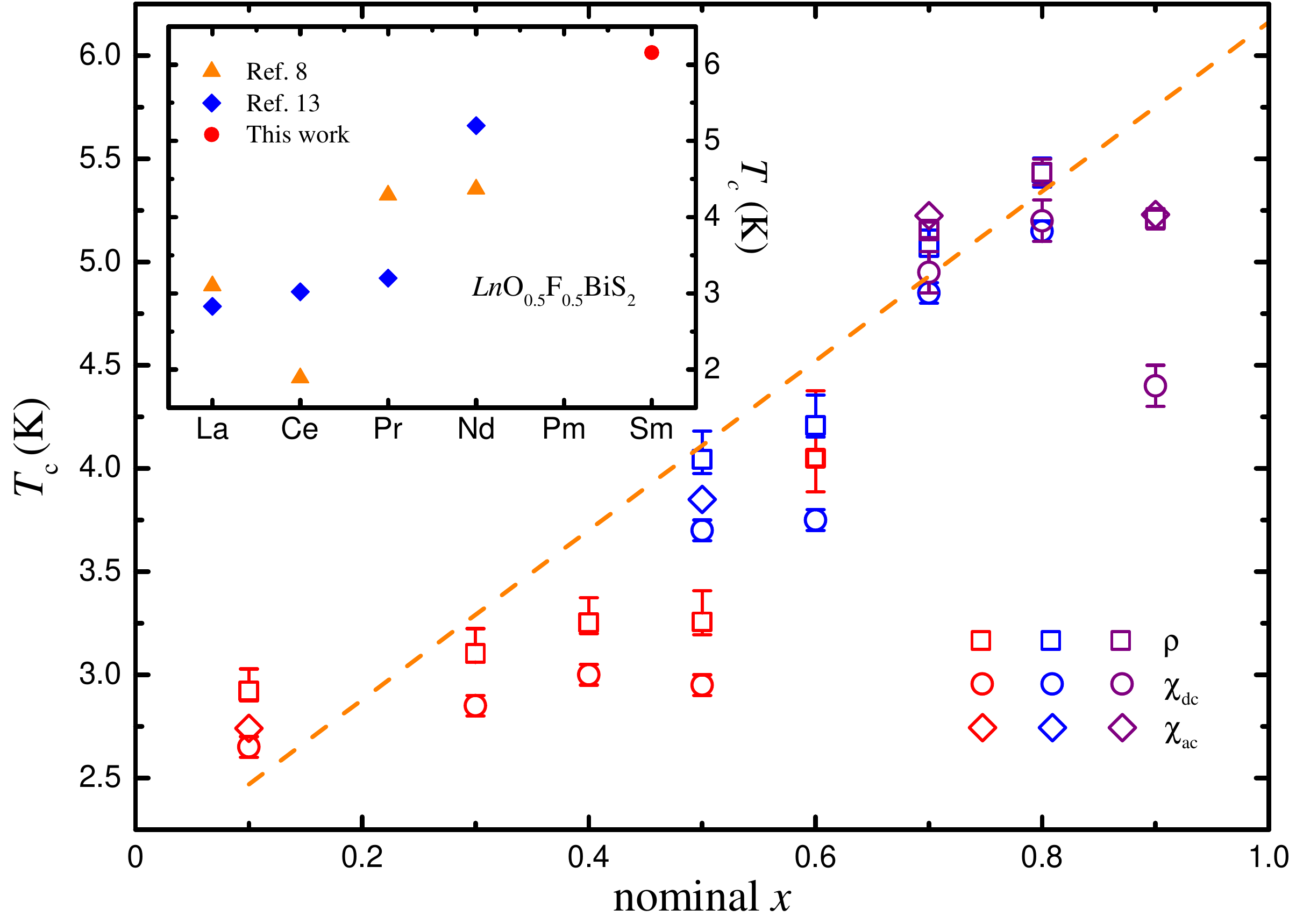}
\caption{(Color online) Superconducting critical temperature \textit{T$_{c}$} vs. nominal Sm concentration $x$ of La$_{1-x}$Sm$_{x}$O$_{0.5}$F$_{0.5}$BiS$_{2}$. Red, blue, and purple symbols represent results for samples annealed at 800$\celsius$, 750$\celsius$, and 710$\celsius$, respectively. The dashed line is a linear fit of \textit{T$_{c, \rho}$} from $x$ = 0.1 to $x$ = 0.8. (Inset) \textit{T$_{c, \rho}$} of \textit{Ln}O$_{0.5}$F$_{0.5}$BiS$_{2}$ compounds reported in Refs. 8 and 13 together with the estimated \textit{T$_{c, \rho}$} = 6.2 K of SmO$_{0.5}$F$_{0.5}$BiS$_{2}$.}
\label{FIG.4.}
\end{figure}

\begin{figure}[t]
\includegraphics[width=8.5cm]{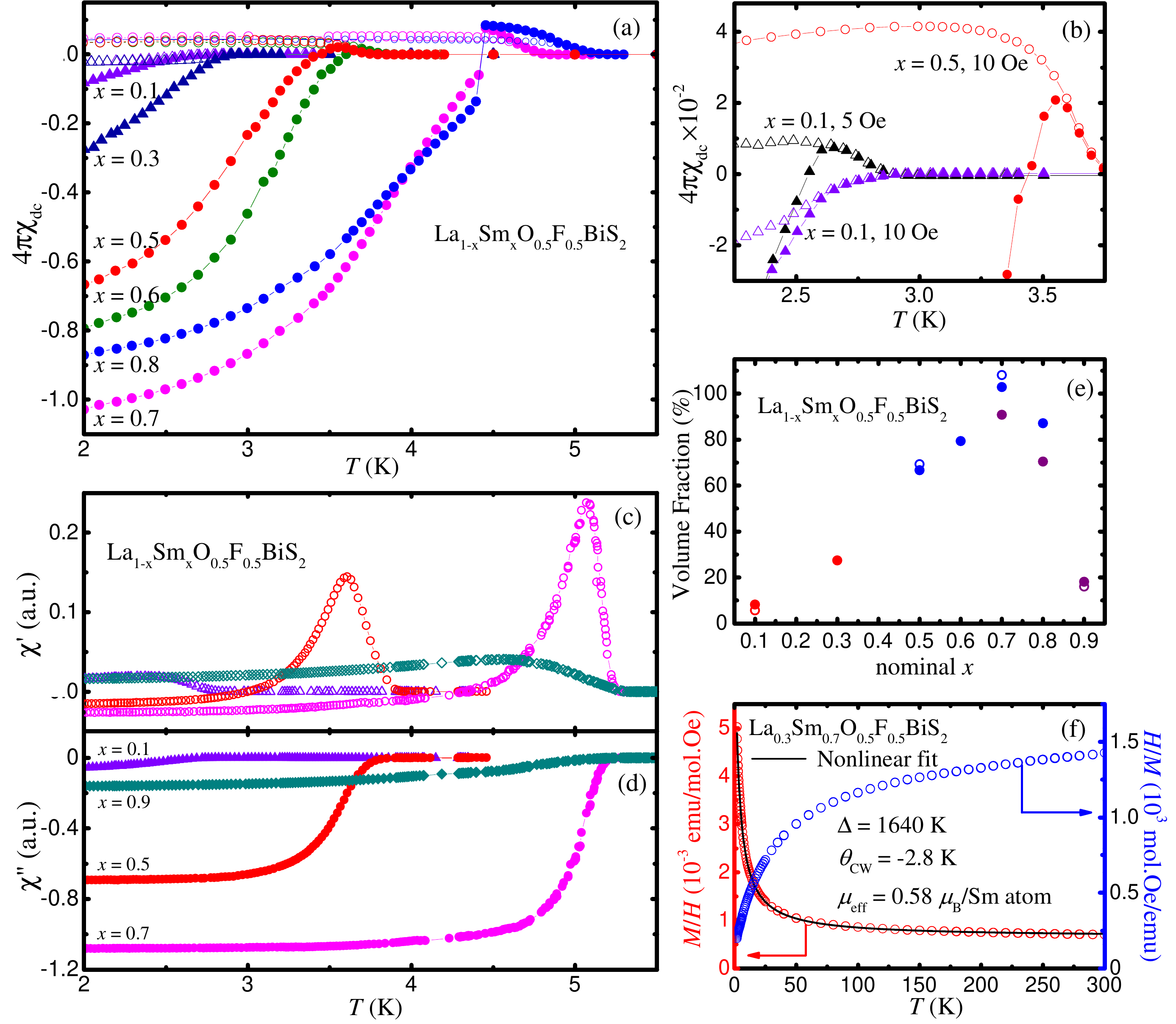}
\caption{(Color online) (a) Zero-field-cooled (ZFC) (filled symbols) and field-cooled (FC) (open symbols) dc magnetic susceptibility data for La$_{1-x}$Sm$_{x}$O$_{0.5}$F$_{0.5}$BiS$_{2}$ in an applied magnetic field of 10 Oe.  (b) Paramagnetic-like behavior of selected samples with $x$ = 0.1 and 0.9 in applied magnetic fields of 5 Oe and 10 Oe, respectively. Magnetic susceptibility data for $x$ = 0.1 in 10 Oe is also plotted for comparison. (c) Real and (d) imaginary part of the ac magnetic susceptibility for selected samples. (e) Evolution of shielding volume fraction with Sm substitution. Open and filled circles correspond to ac and dc susceptibility; red, blue, and purple colored data points represent measurements on samples annealed at 800$\celsius$, 750$\celsius$ and 710$\celsius$, respectively. (f) \textit{M}/\textit{H} and \textit{H}/\textit{M} vs. \textit{T} data in the normal state for La$_{0.3}$Sm$_{0.7}$O$_{0.5}$F$_{0.5}$BiS$_{2}$, measured from 2 to 300 K in an applied magnetic field of 5 kOe. The solid line is a nonlinear fit using Eq. (1).}
\label{FIG.5.}
\end{figure}

 We defined \textit{T$_{c}$} in dc and ac magnetic susceptibility measurements as the temperature at which the ZFC and FC data separate and the point where the imaginary part drops below zero, respectively. The \textit{T$_{c, \chi dc}$} values determined from $\chi_{dc}$ measurements increase monotonically from 2.65 K for $x$ = 0.1 to 5.20 K for $x$ = 0.8 as shown in Fig. 4. Furthermore, dc and ac susceptibility measurements reveal enhanced volume and shielding fractions at 2 K with increasing Sm substitution (Fig. 5(e)), respectively, indicating improvements in the quality of the samples. The optimal volume fraction is obtained at $x$ = 0.7. With further Sm substitution, however, the volume fraction rapidly decreases, coinciding with the appearance of appreciable amount of non-superconducting secondary phases. The fact that the \textit{T$_{c}$} of the sample with $x$ = 0.8 is higher than that of the sample with $x$ = 0.7, which shows the highest volume fraction, implies that superconductivity in La$_{1-x}$Sm$_{x}$O$_{0.5}$F$_{0.5}$BiS$_{2}$ could be further enhanced if samples could be prepared with a higher Sm concentration.

Magnetization \textit{M}, divided by magnetic field \textit{H}, \textit{M}/\textit{H}, for \textit{H} = 5 kOe and \textit{x} = 0.7 is displayed as a function of temperature in Fig. 5(f) (left axis). In addition to distinct non-Curie Weiss behavior (see Fig. 5(f), right axis), there is no evidence for magnetic order down to 2 K. 
Unlike other heavy lanthanides, the energy between the \textit{J} = 5/2 ground state and the \textit{J} = 7/2 first excited state is only 0.12 eV in Sm$^{3+}$ and the Van Vleck term should be considered when modeling the magnetic susceptibility of compounds containing Sm.\cite{Van2,Brain2} Hence, the temperature dependence of the magnetization was fitted by a modified Curie-Weiss law:

  \begin{equation}
  \frac{M}{H}=\frac{N_{A}}{k_{B}}\left[\alpha_{J}\mu_{B}^{2}+\frac{\mu_{eff}^{2}}{3(T-\theta_{CW})}\right],
  \end{equation}
in which $N_{A}$ is Avogadro's number, $k_{B}$ is Boltzmann's constant, $\mu_{B}$ is the Bohr magneton, $\mu_{eff}$ is the effective magnetic moment in Bohr magnetons, and $\theta_{CW}$ is the Curie-Weiss temperature. We define $\alpha_{J}=20/7\Delta$, where $\Delta$ is the energy separation between the $J$ = 5/2 ground state multiplet and the $J$ = 7/2 first excited state multiplet for Sm.\cite{delta} From the best fit of the \textit{M}/\textit{H} data using Eq. (1), values for $\Delta$, $\theta_{CW}$, and $\mu_{eff}$ were found to be 1640 K, -2.8 K, and 0.58 $\mu_{B}$/Sm atom, respectively. The experimental $\Delta$ value is close to the estimated value for free Sm$^{3+}$ ($\sim$1500 K).\cite{Van2} The effective magnetic moment of the samples are considerably smaller than the free Sm$^{3+}$ ion value of 0.845 $\mu_{B}$/Sm atom. Similar low values of $\mu_{eff}$  have been reported in other studies\cite{Brain1,smallmu} and are not evidence for an intermediate valence for Sm; also, an accurate theoretical description of experimental data in Sm systems is complicated by the combined effects of the crystalline electric field (CEF) effects and \textit{J}-mixing.\cite{Brain3}

\begin{figure}[t]
\centering
\includegraphics[width=8.5cm]{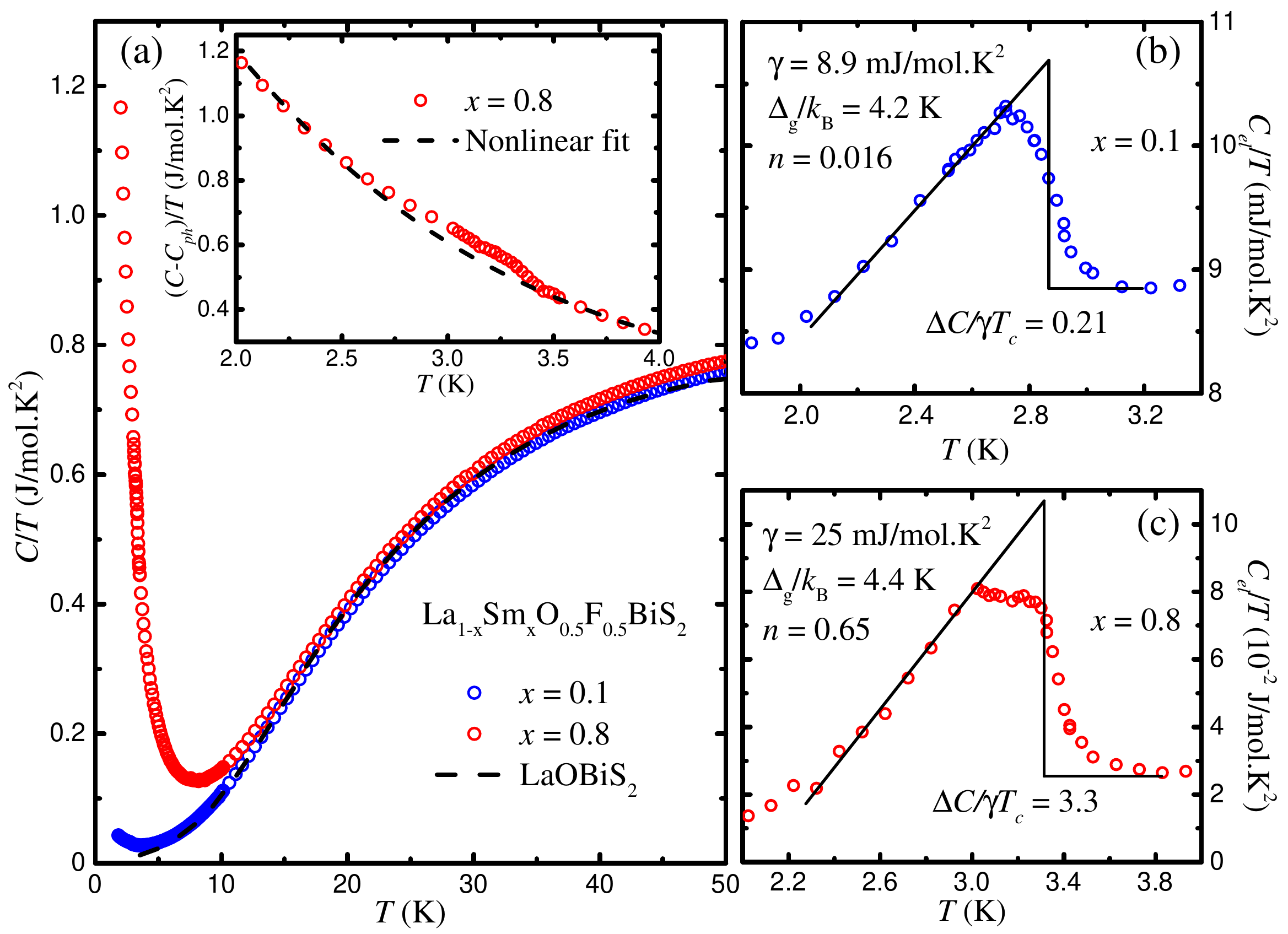}
\caption{(Color online) (a) Specific heat \textit{C} divided by temperature, \textit{C}/\textit{T}, vs. \textit{T} for La$_{1-x}$Sm$_{x}$O$_{0.5}$F$_{0.5}$BiS$_{2}$ with $x$ = 0.1, 0.8 and for LaOBiS$_{2}$. (Inset) (\textit{C}-\textit{C$_{ph}$})/\textit{T} vs. \textit{T}, where \textit{C$_{ph}$} is the lattice contribution, and a fit of the Schottky anomaly contribution for La$_{0.2}$Sm$_{0.8}$O$_{0.5}$F$_{0.5}$BiS$_{2}$ (dashed line). (b) and (c) Electronic contribution \textit{C$_{el}$}/\textit{T} of La$_{1-x}$Sm$_{x}$O$_{0.5}$F$_{0.5}$BiS$_{2}$ with $x$ = 0.1 and $x$ = 0.8, respectively.}
\label{FIG.6.}
\end{figure}
 The results of specific heat \textit{C} measurements for $x$ = 0.1, 0.8, and a nonmagnetic reference compound LaOBiS$_{2}$ are displayed in Fig. 6(a), plotted as \textit{C}/\textit{T} vs. \textit{T}. Above 10 K, the specific heat of the compounds are almost the same, due to similar lattice contributions. The upturns in $C$/$T$ vs. $T$ below 3.7 K and 8.0 K for the samples with $x$ = 0.1 and 0.8, respectively, which overlap with the superconducting transitions, are due to a Schottky contribution (\textit{C$_{Sch}$}) caused by CEF splitting of the $J$ = 5/2 Hund's rule ground state multiplet. Hence, the specific heat of the samples consists of electronic (\textit{C$_{el}$}), phonon (\textit{C$_{ph}$}), and Schottky (\textit{C$_{Sch}$}) contributions. The best fit of the LaOBiS$_{2}$ data below 7 K using \textit{C}(\textit{T}) = \textit{C$_{el}$}(\textit{T}) + \textit{C$_{ph}$}(\textit{T}) = $\gamma$\textit{T} + \textit{A$_{3}$}\textit{T$^{3}$} + \textit{A$_{5}$}\textit{T$^{5}$}, yields the normal-state electronic specific heat coefficient $\gamma$ = 3.32 mJ/mol K$^{2}$ and the coefficients of the phonon contribution \textit{A$_{3}$} = 0.655 mJ/mol K$^{4}$ and \textit{A$_{5}$} = 4.27 $\mu$J/mol K$^{6}$. Representative (\textit{C-C$_{Ph}$})/\textit{T} vs. \textit{T} data for $x$ = 0.8 are shown in the inset of Fig. 6(a). The phonon contribution of the Sm-substituted samples was assumed to be the same as for LaOBiS$_{2}$ and was subtracted from the specific heat. The remaining specific heat data could be fitted with the following expression:
\begin{equation}
C(T)/T=\gamma+nC_{Sch}/T=\gamma+n\frac{R}{T}\frac{\left(\frac{\Delta_{g}}{k_{B}T}\right)^{2}e^{\left(\frac{\Delta_{g}}{k_{B}T}\right)}}{\left[1+e^{\left(\frac{\Delta_{g}}{k_{B}T}\right)}\right]^{2}}.
\end{equation}
The second term in Eq. (2), \textit{n}\textit{C$_{Sch}$}/\textit{T}, represents a Schottky anomaly in which $n$ is the number of Sm atoms per formula unit that contribute to the Schottky anomaly, $\Delta_{g}$ is the splitting between the ground state and the first excited state doublets of the $J$ = 5/2 Hund's rule ground state multiplet, $k_{B}$ is Boltzmann's constant, and \textit{R} is the ideal gas constant. The best fits to the \textit{C}(\textit{T})/\textit{T} for the La$_{1-x}$Sm$_{x}$O$_{0.5}$F$_{0.5}$BiS$_{2}$ samples with $x$ = 0.1 and 0.8 provide very similar $\Delta$$_{g}$ splitting values, but different $\gamma$ values (listed in Figs. 6(b) and (c)). Subtracting both \textit{C$_{ph}$}(\textit{T}) and \textit{C$_{Sch}$}(\textit{T}) from \textit{C}(\textit{T}) data yields the electronic specific heat $C_{el}$(\textit{T}) contribution, revealing a clear feature around \textit{T$_{c}$}, which provides evidence for bulk superconductivity. The \textit{T$_{c}$} values estimated from the entropy conserving constructions are lower than the \textit{T$_{c, \rho}$} and \textit{T$_{c, \chi dc}$} values determined from $\rho$(\textit{T}) and $\chi$$_{dc}$(\textit{T}) measurements, which may suggest some inhomogeneity in the polycrystalline samples. Values of $\Delta$\textit{C}/$\gamma$\textit{T$_{c}$} of 0.21 and 3.3 were extracted from the \textit{C$_{el}$}(\textit{T}) data for $x$ = 0.1 and 0.8, respectively. Given the uncertainties involved in the procedure for extracting the $\Delta$\textit{C}/$\gamma$\textit{T$_{c}$}, these estimates for $\Delta$\textit{C}/$\gamma$\textit{T$_{c}$} are consistent with bulk superconductivity and of the order of magnitude of the BCS value of 1.43. To perform a more reliable quantitative analysis, specific heat data measured below 1.8 K will be needed since the uncertainty in evaluating the Schottky contributions is appreciable because of the limited temperature range measured.  However, the possibility that La$_{1-x}$Sm$_{x}$O$_{0.5}$F$_{0.5}$BiS$_{2}$ exhibits unconventional superconductivity that cannot be explained by BCS theory cannot be ruled out.

\section{SUMMARY}

In summary, the \textit{T$_{c}$} and superconducting volume fraction were found to increase with $x$ in La$_{1-x}$Sm$_{x}$O$_{0.5}$F$_{0.5}$BiS$_{2}$ compounds investigated in the experiments reported herein. The solubility limit of Sm has a large value of $x$ $\sim$ 0.8 in La$_{1-x}$Sm$_{x}$O$_{0.5}$F$_{0.5}$BiS$_{2}$, and a continuous decrease in the $a$ axis and increase in the $c$ axis is observed with increasing $x$. Bulk superconductivity was observed in the samples, according to magnetic susceptibility and specific heat measurements. No evidence of a structural phase transition was found in this study. The results demonstrate that the superconducting critical temperature \textit{T$_{c}$} of tetragonal BiS$_{2}$-based compounds is correlated with the lattice parameters and can be significantly enhanced by Sm substitution. This gives a promising way in further increase the \textit{T$_{c}$} of BiS$_{2}$ based compounds by modifying the blocking layers through the substitution of heavier \textit{Ln} lanthanides (\textit{Ln} = Eu - Tm) or synthesizing the parent \textit{Ln}O$_{1-x}$F$_{x}$BiS$_{2}$ compounds.

\begin{acknowledgements}
Research at UCSD was supported by the U. S. Department of Energy, Office of Basic Energy Sciences, Division of Materials Sciences and Engineering under Grant No. DE-FG02-04-ER46105. Helpful discussions about the MPMS measurement artifacts with N. R. Dilley are gratefully acknowledged.
\end{acknowledgements}

\bibliography{endnote}

\clearpage

\end{document}